\documentclass[10pt,journal]{IEEEtran}

\usepackage{amsmath,amssymb,amsthm,mathtools,bm}
\usepackage{cite}
\usepackage[hidelinks]{hyperref}
\usepackage{microtype}
\usepackage{graphicx}
\usepackage{dsfont}
\allowdisplaybreaks

\newcommand{\I}{\mathds{1}}
\newcommand{\Iv}[1]{\I_{\left[#1\right]}}
\newcommand{\E}{\mathbb{E}}
\newcommand{\Var}{\mathrm{Var}}
\newcommand{\Prb}{\mathbb{P}}
\newcommand{\GED}{\mathrm{GED}}

\newtheorem{proposition}{Proposition}
\newtheorem{lemma}{Lemma}

\newtheorem{remark}{Remark}
\newtheorem{note}{Note}
\title{Typical Solutions of Multi-User Linearly-Decomposable Distributed Computing}

\author{Ali Khalesi  and  Mohammad Reza Deylam Salehi%
\thanks{A.~Khalesi is an \textit{Assistant Professor} at Institut Polytechnique des Sciences Avanc\'ees (IPSA), Paris, France (ali.khalesi@ipsa.fr).
M.\,R.~Deylam Salehi is with the Communication Systems Department, EURECOM, Biot Sophia Antipolis, France (deylam@eurecom.fr).}
}

\begin{document}
\maketitle

\begin{abstract}
We solve, in the typical-case sense, the multi-sender linearly-decomposable distributed computing problem introduced by tessellated distributed computing. We model real-valued encoders/decoders and demand matrices, and assess structural fidelity via a thresholded graph edit distance between the demand support and the two-hop support of the computed product. Our analysis yields: a closed-form second-moment (Frobenius) risk under spike-and-slab ensembles; deterministic links between thresholded GED and norm error; a Gaussian surrogate with sub-exponential tails that exposes explicit recall lines; concentration of GED and operator-norm control; and a compute-capped design with a visible knee. We map the rules to aeronautical and satellite networks.
\end{abstract}


\section{Introduction}
\label{sec:intro}


Multi-user distributed computing (MUDC) increasingly runs at the edge and on airborne or satellite assets (UAVs, HAPS, LEO/GEO constellations), where both \emph{communication} and \emph{computation} budgets are tight, yet users demand low retrieval error. Classical approaches such as coded MapReduce, gradient coding, and coded matrix multiplication provide strong guarantees on norm error and straggler/latency resilience. However, many real failures are \emph{structural}: a requested sub-function is simply unreachable through the server layer, even when the numerical error is small.

The tessellated framework \cite{KhalesiEliaTIT2025} addresses the $K$-user, $N$-server distributed computing scenario, where each user requests a real function that is \emph{linearly-decomposable} over $L$ basis subfunctions. The set of user demands is represented by the coefficient matrix $F\in\mathbb{R}^{K\times L}$. Servers compute subsets of the $L$ subfunctions (captured by the matrix $E$), while users connect to subsets of servers (captured by the matrix $D$). The key system parameters are: (i) the number of users $K$, which dictates demand dimensionality; (ii) the number of servers $N$, which captures available parallelism; and (iii) the number of subfunctions $L$, which reflects computational granularity. Performance is jointly determined by the reconstruction error $\epsilon$ (distortion), the computation load $\gamma$ (fraction of subfunctions per server), and the communication load $\delta$ (fraction of users per server). Under zero-error operation, the problem reduces to finding sparse factors $(D,E)$ such that $DE=F$ under strict per-row/column sparsity constraints. In the lossy case, the framework connects to approximate sparse matrix factorization, with achievable typical distortions characterized through Marchenko–Pastur limits. This triplet $(K,N,L)$ together with $(\gamma,\delta,\epsilon)$ provides the canonical tessellated system model, which our work extends by introducing thresholded graph-edit metrics for structural fidelity.

\paragraph*{\bf From norm error to structural fidelity}
Motivated by deployments where reachability dominates, we complement the tessellated norm-centric picture with a structural metric: \emph{thresholded graph edit distance (GED)} between the \emph{demand support} of $F$ and the \emph{two-hop support} induced by $(D,E)$ after a reliability test (thresholding of the real-valued two-hop strengths). This yields risk decompositions and regime maps in which GED bounds can dominate spectral/sub-multiplicative bounds in ultra-sparse two-hop settings, while $\ell_1/\ell_2$ risk dominates in moderate/dense regimes.

\paragraph*{\bf Positioning with fixed support factorization}
The typical MUDC objective $\min_{D,E}\|DE-F\|_F^2$ under dimensionality and sparsity budgets is NP-hard in general and sits at the intersection of compressed sensing, low-rank approximation, and fixed-support matrix factorization. Unlike compressed sensing (which emphasizes uniqueness and recovery search), our design ties $D$ to the instance $F$ and budgets sparsity where it operationally matters (server fan-in/out), enabling capacity converses/achievability and explicit lossy laws.

\paragraph{Contributions}
We provide a “typical solutions” framework for multi-user \emph{linearly-decomposable} computing with real $D,E,F$ that complements tessellated constructions:
\begin{itemize}
  \item \textbf{Closed-form typical risk:} Exact $\E\|DE{-}F\|_F^2$ under independent spike-and-slab (real-valued) entries and a \emph{thresholded} two-hop reach model $q_\tau$ that captures detection/reliability in real systems (Section~\ref{sec:Set-Th}).
  \item \textbf{Thresholded-GED links:} Deterministic lower/upper connections between thresholded GED and Frobenius/\,$\ell_1$ risk, including when GED bounds beat spectral/Sub-multiplicative bounds (ultra-sparse two-hop) and when they do not (moderate/dense) (Sections~\ref{sec:Det-Ine}–\ref{sec4:norm}).
  \item \textbf{Gaussian surrogate for reachability:} A tractable surrogate $q_\tau$ that exposes \emph{recall lines} and \emph{compute caps}, enabling one-pass operating-point selection (Section~\ref{sec:Des-Eva}).
  \item \textbf{Concentration and Boundary Rule:} Bounded-differences concentration for normalized GED and sub-exponential tail bounds for two-hop strengths under real $D,E$ (Section~\ref{sec:conc}), together with a simple slope test $s = c_{+}(1-{p_F}) - c_{-}{p_F}$ that determines whether to maximize or minimize $q_\tau$ under a recall constraint, thereby unifying budget and SLA tuning.
 
  \item \textbf{Practical guidance:} \emph{Aeronautical} and \emph{satellite} deployments: mapping $\delta,\gamma$ to link and compute caps; accommodating link intermittency, Doppler, and handover; and sizing thresholds to meet per-user recall while minimizing extra-fetch. 
\end{itemize}

\paragraph{Why multi-user linearly-decomposable is the right approach}
It strictly generalizes gradient coding and coded linear transforms by allowing heterogeneous user weights and multi-shot designs while supporting both lossless corner points and lossy typical laws—precisely the dual view (capacity vs.\ distortion) that tessellation champions. Our structural GED analysis fits naturally on top of these real-valued factorizations and explains when “coverage” (reachability) is the true bottleneck and how to buy it with minimal $\delta,\gamma$.

\medskip

\section{Setup and Thresholded GED}
\label{sec:Set-Th}

We consider $K$ users, $N$ servers, and $L$ sub-functions. The demand matrix is $F\in\mathbb{R}^{K\times L}$, the user$\to$server encoding is $D\in\mathbb{R}^{K\times N}$, the server$\to$sub-function decoding is $E\in\mathbb{R}^{N\times L}$, and the arithmetic two-hop map is $A=DE\in\mathbb{R}^{K\times L}$. Entry $A_{k\ell}$ aggregates all two-hop contributions from user $k$ to sub-function $\ell$ through the $N$ servers.

{\bf Budgets and degrees:}
The sparsity budgets are the \emph{communication density} $p_D$ (fraction of nonzeros per row of $D$) and the \emph{compute density} $p_E$ (fraction of nonzeros per column of $E$). Expected degrees are $\E[\mathrm{deg}_{\text{user}}]=Np_D$ and $\E[\mathrm{deg}_{\text{subfn}}]=Np_E$. In networks, $p_D$ maps to scheduling/fan-out on user$\to$server links; $p_E$ maps to compute/I/O activation on server$\to$sub-function links.

{\bf Thresholded supports:} Real systems read out connections through a detector or a reliability test. Let $\tau>0$ be a decision threshold on two-hop strengths and let $\tau_F>0$ (often $\tau_F=\tau/2$) be a demand relevance threshold. Define the induced supports
\begin{align*}
B=\Iv{|A|\ge \tau}\in\{0,1\}^{K\times L},\:\: 
S=\Iv{|F|\ge \tau_F}\in\{0,1\}^{K\times L}.
\end{align*}
With unit edge costs, the thresholded graph edit distance (GED) between $S$ and $B$ is
\begin{IEEEeqnarray}{rCl}\label{eq:gedtau}
\GED_\tau(S,B)
&=& \sum_{k,\ell}\Iv{S_{k\ell}=1,\;B_{k\ell}=0}
\nonumber\\[0.2em]
&& {}+\sum_{k,\ell}\Iv{S_{k\ell}=0,\;B_{k\ell}=1}.
\end{IEEEeqnarray}
It counts \emph{misses} (false negatives) and \emph{extras} (false positives) after thresholding. We also track the Frobenius error $\|A-F\|_F$ and the weighted edit (entrywise $\ell_1$) $\|A-F\|_1$.

{\bf Spike-and-slab ensemble:} To capture sparsity with real weights, we adopt independent spike-and-slab entries:
\begin{IEEEeqnarray}{rCl}
D_{kn}&\sim&(1-p_D)\,\delta_0 \;+\; p_D\,\mathcal{N}(0,v_D),\\
E_{n\ell}&\sim&(1-p_E)\,\delta_0 \;+\; p_E\,\mathcal{N}(0,v_E),
\end{IEEEeqnarray}
and similarly $F_{k\ell}\sim (1-p_F)\delta_0 + p_F\,\mathcal{N}(0,v_F)$. Entries are independent across matrices. Under zero-mean slabs,
\begin{IEEEeqnarray}{rCl}
A_{k\ell}&=&\sum_{n=1}^N D_{kn}E_{n\ell},\\
\E[A_{k\ell}]&=&0,\qquad
\Var(A_{k\ell})=N\,p_Dp_E\,v_Dv_E.
\end{IEEEeqnarray}
Thus $A_{k\ell}$ is sub-exponential; by CLT it is well-approximated as Gaussian near the operating point, with standard deviation $\sigma_A=\sqrt{N p_Dp_E v_Dv_E}$.

{\bf Reachability, recall, and precision:}
Let $q_\tau=\Prb(|A_{k\ell}|\ge\tau)$ be the \emph{two-hop reach probability} at threshold $\tau$. Using the Gaussian surrogate,
\begin{IEEEeqnarray}{rCl}
\label{eq:q_surr}
q_\tau(p_D,p_E)
&\approx& 2\,Q\!\left(\frac{\tau}{\sigma_A}\right),
\end{IEEEeqnarray}
where $Q$ is the standard normal tail. The demand-support probability is
\begin{IEEEeqnarray}{rCl}
p_S
&=& \Prb(|F_{k\ell}|\ge \tau_F)=(1-p_F)\cdot 0 \;+\; p_F\cdot 2Q\!\left(\frac{\tau_F}{\sqrt{v_F}}\right)
\end{IEEEeqnarray}
With asymmetric costs $(c_-,c_+)$ for misses/extras, the expected per-edge cost is the affine function
\begin{IEEEeqnarray}{rCl}
\phi_\tau(q_\tau)
&=& c_-\,p_S\,(1-q_\tau)\;+\;c_+\,(1-p_S)\,q_\tau.
\end{IEEEeqnarray}
Its slope $s=c_+(1-p_S)-c_-p_S$ determines whether to raise or lower $q_\tau$ (boundary rule).

{\bf Threshold selection and invariances:}
Scaling $(D,E)\mapsto (\alpha D,\alpha^{-1}E)$ leaves $A$ unchanged; hence $\tau$ should be interpreted relative to scale, e.g., via the \emph{dimensionless} threshold $\tilde\tau=\tau/\sigma_A$. Larger $\tau$ reduces extras (better precision) but increases misses (worse recall). In practice:
\begin{itemize}
\item choose $\tau$ from an ROC target (application SNR or minimum precision);
\item set $\tau_F$ as the minimum “useful” magnitude in $F$ (e.g., coefficient quantization or SLA threshold);
\item enforce a recall requirement $\mathbb{P}(B{=}1\,|\,S{=}1)\ge \rho$, which is met by $q_\tau\ge \rho$ under independence.
\end{itemize}

\paragraph{Why thresholded GED}
In ultra-sparse two-hop regimes (small $p_Dp_E$ or small $N$), entries of $A$ concentrate near $0$ or a single-path contribution. Then the structural decision “is the edge \emph{there}?” dominates, and $\GED_\tau$ tightly tracks the operational loss. As multiplicities grow (moderate/dense two-hop), magnitude errors matter and $\|A-F\|_1$/$\|A-F\|_F$ become more faithful; our later bounds translate performance between these metrics.

\paragraph{Network mapping}
On real links, $p_D$ reflects per-slot scheduling or beam fan-out; $p_E$ reflects server occupancy or accelerator duty cycle; $\tau$ encodes the demod/decoding or estimation reliability required for an edge to be “usable.” The triplet $(p_D,p_E,\tau)$ therefore exposes a concrete \emph{communication–computation–retrieval} design plane, and $\GED_\tau$ measures the structural gap induced by design choices.

\section{Deterministic Inequalities}
\label{sec:Det-Ine}

We compile Frobenius- and operator-norm inequalities that hold for \emph{any} real $D,E,F$, and discuss the conditions under which they are informative for design.

\begin{note}
[Reverse/forward triangle]
\label{note:T1}
\begin{IEEEeqnarray}{c}
\Big|\|DE\|_F-\|F\|_F\Big|
\le\|DE{-}F\|_F \le \|DE\|_F+\|F\|_F.
\end{IEEEeqnarray}
\end{note}
\begin{proof}
Apply the reverse and forward triangle inequalities with $X{=}DE$, $Y{=}F$ in the Hilbert space $(\mathbb{R}^{K\times L},\langle\cdot,\cdot\rangle_F)$, where $\langle X,Y\rangle_F=\mathrm{Tr}(X^\top Y)$.
\end{proof}

\noindent\textbf{Equality/tightness:}
The \emph{upper} bound becomes equality when $DE$ and $F$ are positively colinear in Frobenius inner product, i.e., $\langle DE,F\rangle_F=\|DE\|_F\|F\|_F$. The \emph{lower} (reverse) bound becomes equality when they are negatively colinear. If $\langle DE,F\rangle_F=0$ (Frobenius orthogonality), then
\begin{IEEEeqnarray}{rCl}
\|DE{-}F\|_F^2
&=& \|DE\|_F^2+\|F\|_F^2,
\end{IEEEeqnarray}
which is a Pythagorean case useful for sanity checks.

\begin{remark}[Sub-multiplicative upper bounds]
\label{rem:T2}
\begin{IEEEeqnarray*}{rCl}
 \|DE{-}F\|_F
&\le&\|DE\|_F+\|F\|_F
\le \|D\|_F\|E\|_2+\|F\|_F,\\[0.2em]
\|DE{-}F\|_F
&\le& \|DE\|_F+\|F\|_F
\le \|D\|_2\|E\|_F+\|F\|_F.
\end{IEEEeqnarray*}
\end{remark}
\begin{proof}
The first inequality is Note~\ref{note:T1}. For the second step use the standard submultiplicativity
$\|AB\|_F\le \|A\|_F\|B\|_2$ and symmetrically $\|AB\|_F\le \|A\|_2\|B\|_F$.
\end{proof}

\noindent\textit{When is Remark~\ref{rem:T2} informative?}
The bound $\|DE\|_F\le \|D\|_F\|E\|_2$ is \emph{tight} when $E$ acts nearly as an isometry on the dominant right singular subspace of $D$ (e.g., $E$ has a large top singular value aligned with $D$'s top right singular vectors). The symmetric form is tight when $D$ has a large top singular value aligned with $E$'s top left singular vectors. In practice:

\begin{itemize}
\item If $E$ contains high\mbox{-}degree “hubs” (large operator norm), $\|E\|_2$ drives the upper bound and warns of possible overestimation by purely spectral arguments (GED bounds may be better in ultra\mbox{-}sparse two\mbox{-}hop).
\item If both $\|D\|_2$ and $\|E\|_2$ are modest but $\|D\|_F,\|E\|_F$ are large, the weaker $\|DE\|_F\le \|D\|_F\|E\|_F$ (implied by $\|E\|_2\le \|E\|_F$) can still be numerically useful as a coarse ceiling.
\end{itemize}

\begin{proposition}[Singular\mbox{-}value lower bounds]\label{prop:T3}
\begin{IEEEeqnarray}{rCl}
\|DE{-}F\|_F
&\ge& \big|\,\|D\|_F\,\sigma_{\min}(E)-\|F\|_F\,\big|,\\[0.2em]
\|DE{-}F\|_F
&\ge& \big|\,\sigma_{\min}(D)\,\|E\|_F-\|F\|_F\,\big|.
\end{IEEEeqnarray}
\end{proposition}
\begin{proof}
We have $\|DE\|_F\ge \sigma_{\min}(E)\|D\|_F$ because
\begin{IEEEeqnarray}{rCl}
\|DE\|_F^2
&=& \mathrm{Tr}(E^\top D^\top D E)
\;\ge\; \sigma_{\min}(E)^2\,\mathrm{Tr}(D^\top D) \nonumber\\
&=& \sigma_{\min}(E)^2\,\|D\|_F^2,
\end{IEEEeqnarray}
using the Courant–Fischer characterization on $E^\top D^\top D E$. Combine with the reverse triangle bound in Note~\ref{note:T1}. The symmetric inequality follows by swapping  $D$ and $E$.
\end{proof}

\noindent\textbf{Meaning for design:}
Prop.~\ref{prop:T3} says $DE$ cannot be too small in Frobenius norm if either factor is well\mbox{-}conditioned (large $\sigma_{\min}$) while the other carries energy. Thus even if $F$ is small, strong conditioning in $E$ (or $D$) forces a nontrivial lower bound on the error unless the other factor is correspondingly small. This is a warning against \emph{over\mbox{-}activating} well\mbox{-}conditioned blocks when the target $F$ is structurally sparse or low power.

\paragraph*{A quick orthogonality corollary.}
If $DE$ and $F$ are orthogonal, then by the proof of Note~\ref{note:T1}
\begin{IEEEeqnarray}{rCl}
\|DE{-}F\|_F
&=\sqrt{\|DE\|_F^2+\|F\|_F^2} 
\;\ge\; \nonumber \\ &\quad\big|\,\|D\|_F\,\sigma_{\min}(E)-\|F\|_F\,\big|,
\end{IEEEeqnarray}
which shows the singular\mbox{-}value lower bound is often close to exact in “energy\mbox{-}separated’’ regimes (useful when $F$ focuses on entries not energized by $DE$).

\paragraph*{Link to thresholded GED}
Upper bounds like Remark~\ref{rem:T2} can be pessimistic in ultra\mbox{-}sparse two\mbox{-}hop regimes because $\|E\|_2$ is driven by hubs that do not change \emph{support}. In such cases, the structural bounds $\GED_\tau$ (and its links to $\|A{-}F\|_F$ through threshold factors) give sharper guidance for communication/compute budgeting; see Sections on \emph{Norm Risk and Thresholded GED} and \emph{Reach Probability, Recall Lines, and Cap}.

\section{Norm Risk and Thresholded GED}
\label{sec4:norm}
This section connects the squared Frobenius risk to thresholded GED via simple, interpretable decompositions. Throughout, $A{=}DE$, and entries follow the independent spike\mbox{--}and\mbox{--}slab model with zero\mbox{-}mean Gaussian slabs unless stated otherwise.

\begin{lemma}[Second moment of $\|DE{-}F\|_F$]\label{lem:realE1}
Under independence and spike\mbox{--}and\mbox{--}slab with variances $(v_D,v_E,v_F)$ and densities $(p_D,p_E,p_F)$, for any slab mean $\mu_F{=}\E[F_{k\ell}]$,
\begin{IEEEeqnarray}{rCl}
\E\|DE{-}F\|_F^2
&=& KL\Big( Np_Dp_Ev_Dv_E + p_F v_F \nonumber\\
&&\qquad\;\;+\; \big(Np_Dp_Ev_Dv_E - p_F\mu_F\big)^2 \Big).
\label{eq:E_DE_F_expand}
\end{IEEEeqnarray}
With zero\mbox{-}mean slabs for $F$ ($\mu_F{=}0$), this reduces to
\begin{IEEEeqnarray}{rCl}
\E\|DE{-}F\|_F^2
&=& KL\Big( Np_Dp_Ev_Dv_E + p_F v_F \nonumber\\
&&\qquad\;\;+\; (Np_Dp_Ev_Dv_E)^2 \Big).
\label{eq:E_DE_F_zeromean}
\end{IEEEeqnarray}
\end{lemma}

\begin{proof}
For fixed $(k,\ell)$, $A_{k\ell}{=}\sum_{n=1}^N D_{kn}E_{n\ell}$. Independence and zero\mbox{-}mean slabs for $D,E$ give
$\E[A_{k\ell}]{=}0$ and
$\Var(A_{k\ell}){=}\sum_n \E[D_{kn}^2]\E[E_{n\ell}^2]
{=} N\,p_Dv_D\,p_Ev_E$.
Then, using
$\E[(A_{k\ell}{-}F_{k\ell})^2]
{=}\Var(A_{k\ell}){+}\Var(F_{k\ell}){+}(\E [A_{k\ell}]{-}\E [F_{k\ell}])^2$,
sum over $KL$ entries to obtain \eqref{eq:E_DE_F_expand}. If $\mu_F{=}0$ the third term becomes $(Np_Dp_Ev_Dv_E)^2$.
\end{proof}

\noindent\textbf{Interpretation:}
The three terms in \eqref{eq:E_DE_F_expand} have clear roles:
(i) $Np_Dp_Ev_Dv_E$ is the two\mbox{-}hop \emph{variance} induced by $DE$ (communication\,$\times$\,compute budget);
(ii) $p_F v_F$ is the intrinsic \emph{variance} of the demand;
(iii) $(Np_Dp_Ev_Dv_E - p_F\mu_F)^2$ is a \emph{bias} term capturing mean\mbox{-}energy mismatch between $A$ and $F$.
In the zero\mbox{-}mean case \eqref{eq:E_DE_F_zeromean}, the “bias” reduces to the square of the two\mbox{-}hop energy.

\begin{lemma}[Thresholded GED vs.\ Norm]\label{lem:realG1}
Let $B{=}\Iv{|DE|\ge \tau}$ and $S{=}\Iv{|F|\ge \tau_F}$. Then entrywise
\begin{IEEEeqnarray}{c}
\Big(|DE|-|F|\Big)^2 \;\ge\; \min\{\tau,\tau_F\}^2 \;\Iv{B\ne S},
\end{IEEEeqnarray}
so
\begin{IEEEeqnarray}{rCl}
\GED_\tau(S,B)
&\le& \frac{\|\,|DE|-|F|\,\|_F^2}{\min\{\tau,\tau_F\}^2}
\;\le\; \frac{\|DE{-}F\|_F^2}{\min\{\tau,\tau_F\}^2}.
\label{eq:GED_norm_link}
\end{IEEEeqnarray}
\end{lemma}

\begin{proof}
If $B\neq S$, one of $|DE|$ or $|F|$ is $\ge$ its threshold while the other is $<\,$its threshold; hence
$\big||DE|-|F|\big|\ge \min\{\tau,\tau_F\}$.
Summing squares across entries yields the first inequality; the second follows from $||X|-|Y||\le \|X-Y\|$ entrywise.
\end{proof}

\noindent\textbf{Normalization remark:}
Many designs absorb the $\min\{\tau,\tau_F\}^2$ factor into unit costs $(c_-,c_+)$ or fix $\tau_F{=}\tau$, recovering the simplified form
$\GED_\tau(S,B)\le \|\,|DE|-|F|\,\|_F^2 \le \|DE{-}F\|_F^2$ used for quick comparisons.

\begin{lemma}[Weighted edit]\label{lem:realG2}
For any matrix $X$,
\begin{IEEEeqnarray}{c}
\frac{\|X\|_1}{\sqrt{KL}} \;\le\; \|X\|_F \;\le\; \|X\|_1 \ ,\\
\quad \quad
\frac{\|DE{-}F\|_1}{\sqrt{KL}} \le \|DE{-}F\|_F \le \|DE{-}F\|_1.
\end{IEEEeqnarray}
\end{lemma}

\begin{proof}
Standard $\ell_1$–$\ell_2$ norm relations in finite dimensions applied entrywise.
\end{proof}

\medskip
\noindent\textit{Why these links matter?}
\begin{itemize}
\item \emph{Sparse two\mbox{-}hop}: Most $A_{k\ell}$ are near zero or come from a single path; thresholding is the decision of record. Then $\GED_\tau$ is a tight proxy, and \eqref{eq:GED_norm_link} lower\mbox{-}bounds the norm risk up to a fixed (threshold) factor.
\item \emph{Moderate/dense two-hop}: Multiplicities $S_{k\ell}\!\ge\!2$ are common; magnitudes matter. The $\ell_1$ risk is more faithful than Hamming, while $\|DE{-}F\|_F$ captures energy mismatch as in Lemma~\ref{lem:realE1}.
\item \emph{Design dial}: The slope of the expected GED cost
$\phi_\tau(q){=}c_-p_S(1{-}q){+}c_+(1{-}p_S)q$ decides whether to push $q$ up or down. Because $q$ is monotone in $(p_D,p_E)$, one obtains a one\mbox{-}pass policy along recall lines until a compute cap or network constraint binds (see (\ref{eq:q_surr})). 
\end{itemize}

\section{Design Map, Evaluation, and Domain Guidance}
\label{sec:Des-Eva}
\textbf{Boundary rule:} The expected per-edge GED $\phi_\tau(q)$ is affine in $q$ with slope
$s=c_+(1-p_F)-c_-p_F$. If $s{<}0$ (misses costlier), \emph{maximize} $q$ (push $p_D$ to caps, raise $p_E$ up to the recall line/cap). If $s{>}0$ (extras costlier), \emph{minimize} $q$ subject to the recall constraint.

\begin{figure}[t]
\centering
\includegraphics[width=0.9\columnwidth]{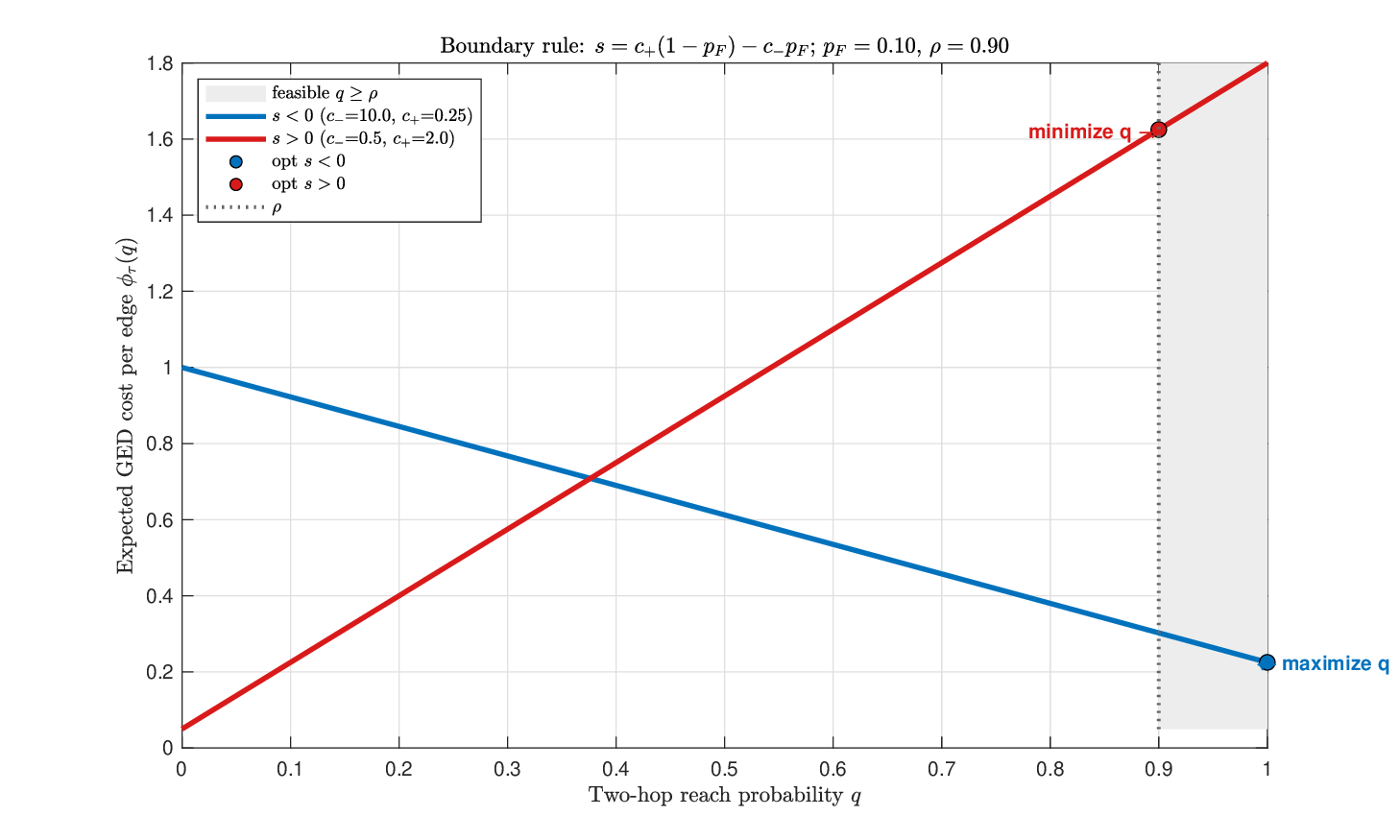}
\caption{Boundary rule: if $s{<}0$ maximize $q$; if $s{>}0$ minimize $q$ subject to $q\!\ge\!\rho$.}
\label{fig:slope}
\end{figure}

\textbf{Setup:} We consider a distributed computation setting with $K=64$ users, $N=64$ servers, and $L=800$ sub-functions, where the system parameters are $p_F=0.10$, $(c_-,c_+){=}(10,0.25)$, $v_D{=}v_E{=}0.50$, $\tau{=}0.10$, $\tau_F=\tau/2$, $p_E^{\mathrm{cap}}{=}0.20$, and $\rho\in\{0.80,0.90,0.95\}$. 

\begin{figure}[t]
\centering
\includegraphics[width=0.9\columnwidth]{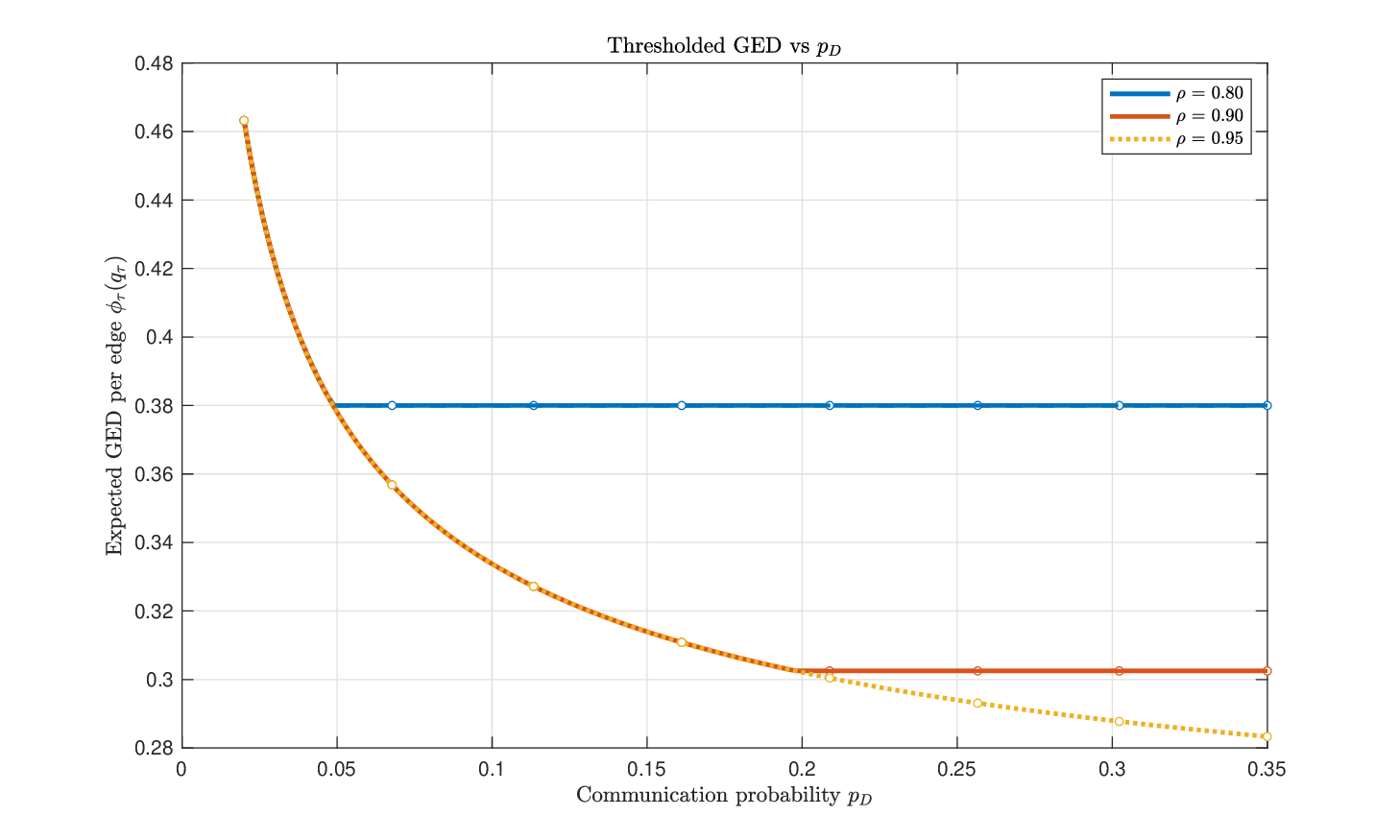}
\caption{Thresholded GED vs.\ $p_D$ (solid: cap-binding; dashed: recall-binding). Circles mark the knee where $p_E^{\min}(p_D,\rho){=}p_E^{\mathrm{cap}}$.}
\label{fig:ged_pd}
\end{figure}

\begin{figure}[t]
\centering
\includegraphics[width=0.9\columnwidth]{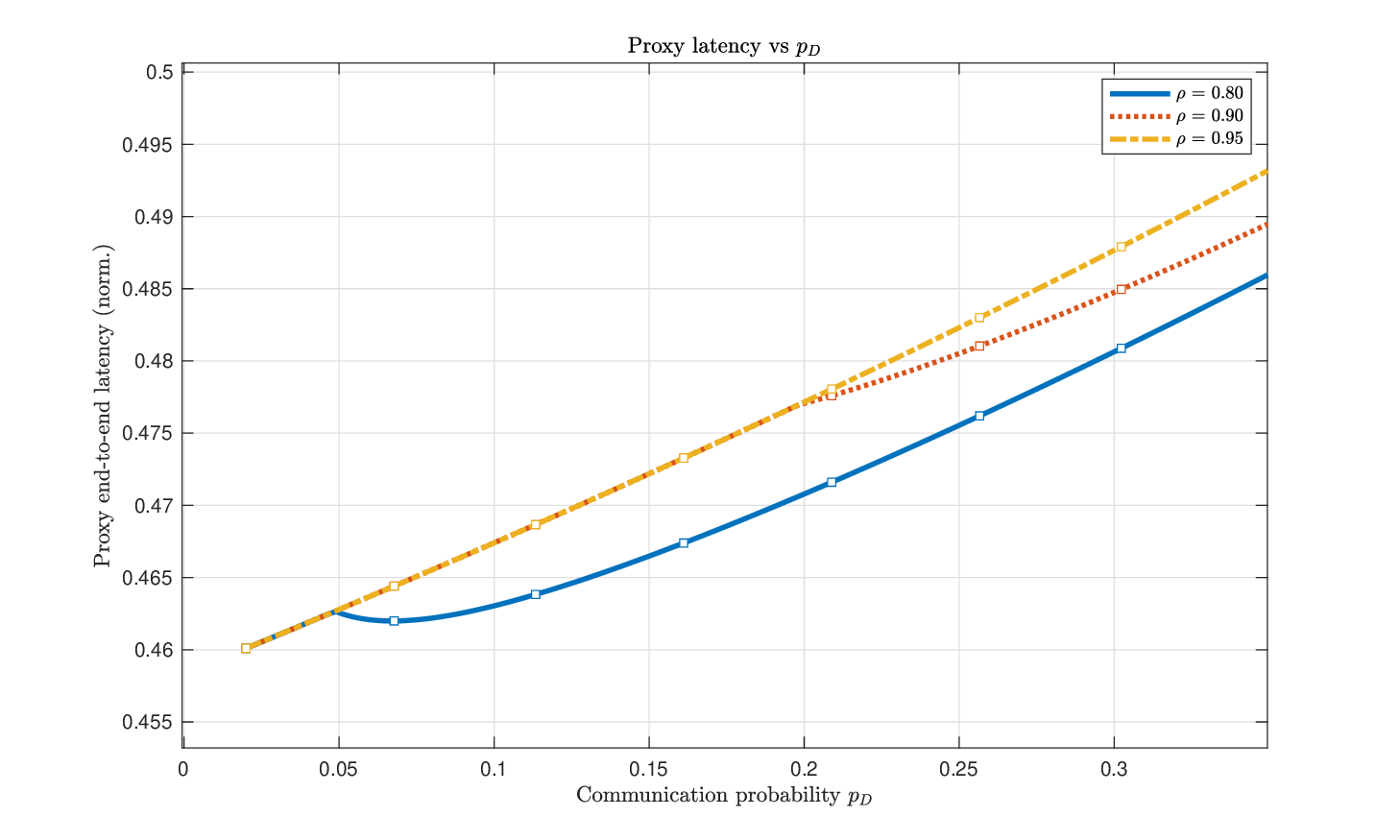}
\caption{Proxy latency vs.\ $p_D$ under the same settings.}
\label{fig:latency_pd}
\end{figure}

\textbf{Aeronautical networks:} In A2G/A2A, $p_D$ maps to sector and beam scheduling and $p_E$ to onboard/edge compute activation. If $s{<}0$, increase $p_D$ (coverage first), then lift $p_E$ along the recall line; if $s{>}0$, restrain $p_D$ until recall binds to avoid extra fetch/traffic.

\textbf{Satellite systems:} In multi-beam LEO/GEO, $p_D$ controls feeder fan-out and $p_E$ the payload occupancy. The knee gives the smallest $p_D$ that achieves the target recall at the compute cap, guiding beam hopping, task placement, and SLA tuning.

Fig.~\ref{fig:slope} illustrates the boundary rule, showing how the slope $s$ dictates whether to maximize or minimize $q$. Fig.~\ref{fig:ged_pd}~plots the thresholded GED versus $p_D$, with the knee marking the smallest $p_D$ that achieves the recall target at the compute cap. Fig.~\ref{fig:latency_pd} reports the corresponding proxy latency trends under the same settings.

\section{Conclusion}
\label{sec:conc}

We provided a typical-case solution for real-valued MUDC with thresholded GED: an affine objective in the reach probability, explicit recall lines, compute caps with visible knees, and concentration. Upper bounds from sub-multiplicativity dominate in dense regimes; GED-based bounds are tight and informative in sparse two-hop regimes. The slope-driven boundary rule offers a deployable design knob for aeronautical and satellite networks.
\vspace{-10pt}
\bibliographystyle{IEEEtran}

\vspace{-0.5 cm}


\end{document}